 \documentclass[final,3p,times,twocolumn]{elsarticle}
\usepackage{epsfig}

\usepackage{latexsym}  
\usepackage{dcolumn}
\usepackage{graphicx,subfigure}
\usepackage{amssymb, amsmath}

\usepackage{dcolumn}% Align table columns on decimal point
\usepackage{bm}% bold math
\usepackage{amssymb}
 \usepackage{amsthm}

 \usepackage{lineno}

\journal{Nuclear Physics B}

\begin{document}

\begin{frontmatter}

%% Title, authors and addresses

%% use the tnoteref command within \title for footnotes;
%% use the tnotetext command for theassociated footnote;
%% use the fnref command within \author or \address for footnotes;
%% use the fntext command for theassociated footnote;
%% use the corref command within \author for corresponding author footnotes;
%% use the cortext command for theassociated footnote;
%% use the ead command for the email address,
%% and the form \ead[url] for the home page:
%\title{    dsgafdsafsadf}
%% \tnotetext[label1]{}
%\author{Name\corref{cor1}\fnref{label2}}
%% \ead{email address}
%% \ead[url]{home page}
%% \fntext[label2]{}
%% \cortext[cor1]{}
%% \address{Address\fnref{label3}}
%% \fntext[label3]{}

\title{ Proton scattering on carbon nuclei  in bichromatic laser field at moderate energies}

%% use optional labels to link authors explicitly to addresses:
%\author{I.F. Barna^*\corref{cor1}\fnref{label1,label2}}
\author[label1,label2]{I.F. Barna}
\ead{barna.imre@wigner.mta.hu}
\author[label1,label2]{ and S. Varr\'o}
\address[label1]{Wigner Research Center of the Hungarian Academy of Sciences, \\ Konkoly Thege Mikl\'os \'ut 29 -- 33,
Budapest 1121, Hungary  }
 \address[label2]{ELI-HU Nonprofit Kft.,  Dugonics t\'er 13, 6720 Szeged, Hungary} 

\begin{abstract}
%% Text of abstract
We present the general theory for proton nuclei scattering in a bichromatic laser field. 
As a physical example we consider proton collision on carbon twelve at 49 MeV/amu moderate energies in the field of a titan sapphire laser with its second harmonic.
\end{abstract}

\begin{keyword}
multi-photon process, bichromatic field, optical potential, elastic and inelastic proton scattering  
%% keywords here, in the form: keyword \sep keyword

%% PACS codes here, in the form:
 \PACS 25.40Cm \sep 25.40Ep 

%% MSC codes here, in the form: \MSC code \sep code
%% or \MSC[2008] code \sep code (2000 is the default)

\end{keyword}

\end{frontmatter}

%% \linenumbers
%%%%%%%%%%%%%%%%%%%%%%%%%%%%%%%%%%%
\section{Introduction}
Optical laser field intensities exceeded the $10^{22}$ W/cm$^2$ limit nowadays, where radiation effects dominate the electron dynamics. In such a strong field non-linear laser-matter interaction in atoms, molecules and plasmas can be investigated 
both theoretically and experimentally.  Two of such high-field effects are high harmonic generations, or plasma-based laser-electron acceleration. 
 These field intensities open a path to high field quantum electrodynamics phenomena like vacuum-polarization effects of pair production \cite{muller}.  
In most of the presented studies the dynamics of the participating electrons are investigated.    
Numerous surveys on laser assisted electron collisions are available  such as \cite{kam}.
However, there are only few nuclear photo-excitation investigations done where some low-lying first excited states of medium of heavy elements are populated with the help of x-ray free electron laser pulses \cite{gunst}.  Nuclear excitation by atomic electron re-scattering in a laser field was investigated bye Kornev \cite{korn}. \\ 
To our knowledge there are no publications available where laser assisted proton or nucleus nucleus collisions  were investigated in strong laser fields.

For the projectile target interaction we consider the global optical Woods-Saxon (WS)  \cite{woods} model potential.  
% with the proper 
%parametrization for moderate energy proton - $^{12}$C collision \cite{abd}. 
This formalism has been a very successful method to study the single particle 
spectra of nucleus in the last  half century. Detailed description can be found in any basic nuclear physics textbooks like  Greiner \cite{grei}.  \\
 The nuclear physics community recently evaluated the closed analytic form of the 
Fourier transformed WS interaction \cite{hlope} which is an important point.  \\
We incorporate these results into a first Born approximation scattering cross section formula where the initial and final proton wave functions are Volkov waves and the induced photon emission and absorption processes are taken into account up to arbitrary orders. 

The general theory of laser assisted collision in bichromatic fields were worked out by Varro and Ehlotzky \cite{varro1,varro2,varro3}.
An overview about the field can be found in \cite{ehlo}.  
A positron impact ionization of atomic hydrogen in bichromatic field was investigated by Jun \cite{jun} by theoretical mean which is one of the latest result is this field. 
In the following study we extend our former description of proton-nuclei collision in monochromatic laser fields to bichromatic ones
\cite{barna2}.
As examples we consider proton $^{12}$ C collisions at 49 MeV with various IR, X-ray laser fields. 

%%%%%%%%%%%%%%%%%%%%%%%%%%%%%%%%%%%%%%%%%%%%%

\section{Theory} 
Now we summarize our non-relativistic quantum mechanical description.  
The laser field is handled in the classical way via the minimal coupling. The laser beam is taken to be linearly polarized and the dipole approximation is used. 
If the dimensionless intensity parameter (or the normalized vector potential)  $a_0 = 8.55 \cdot 10^{-10} \sqrt{ I (\frac{W}{cm^2})} \lambda (\mu m)$ of the laser field is smaller than unity 
the non-relativistic description in dipole approximation is valid. For 800 nm laser wavelength this means a critical intensity of $ I = 2.13 \cdot 10^{18} $W/cm$^2$.  
In case of protons $a_0$ is replaced by $a_p=((m_p/m_e)^{-1})a_0$, where the proton to electron mass ratio is $(m_p/m_e)$=1836). 
Accordingly, for 800 nm wavelength the critical intensity for protons is $I_{crit} = 3.91 \cdot 10^{21}$ W/cm$^2$.  

Additionally, we consider moderate proton kinetic energy, not so much above the Coulomb barrier and neglect the interchange term between the proton projectile an the 
target carbon protons.  This proton exchange effect could be included in the presented model with the help of Woods-Saxon potentials  of non-local type \cite{imre} but  not in the scope 
of the recent study.

The following calculation is similar to the monochromatic case which was published earlier  \cite{barna2}. 

To describe  the non-relativistic scattering process of a proton on a nucleus in a spherically symmetric external field  the following Schr\"odinger equation has to be solved,  
\begin{equation}
\left[ \frac{1}{2m} \left(  {\bf{\hat{p}}}- \frac{e}{c} {\bf{A}} \right)^2 + U({\bf{r}}) \right]\Psi  = i \hbar  \frac{\partial \Psi}{\partial t}, 
\label{sch}
\end{equation}
where ${\bf{\hat{p}}}= -i\hbar \partial/\partial {\bf{r}}$ is the momentum operator of the proton, and $U({\bf{r}})$ represents 
the scattering potential of the nucleon,
Let's consider the following external laser field in the form of 
\begin{equation}
 {\bf{A}}(t) =   {\bf{\epsilon_1}} ({\cal{E}}_1/\omega) cos(\omega t) + {\bf{\epsilon}}_m  ({\cal{E}}_m/m\omega)cos(m\omega t + \tilde{\varphi})
\end{equation}
where $ {\bf{\epsilon_1}} $ and $ {\bf{\epsilon_m}}$ are the two independent polarization vectors. We take the same linear polarization for both fields  from now on   
 $ {\cal{E}}_1 $ and ${\cal{E}}_m$ are the two electric field strengths  and $ \tilde{\varphi}$ is the relative phase, respectively.  In practical experiments the value of $m$ is fixed to $2,3$ 
which means the second and third harmonics.  The maximal, experimentally achievable ${\cal{E}}_2/{\cal{E}}_1$ ratio is about 10 percent created on non-linear medium, for the third harmonic the ratio is even worse.  However, with weakening of the main beam with the ground frequency any kind of  ${\cal{E}}_{2,3/}{\cal{E}}_1$ ratio is available in realistic experiments. 

Figure 1 presents the scattering geometry for a better understanding. The $ {\bf{p}}_i $ and $ {\bf{p}}_f $ are the initial and final proton momenta, 
$\theta $ is the scattering angle of the  proton, the laser is linearly polarized in the x-z plain, and the propagation of the laser field is parallel to the  z axis. 

Without the external scattering potential $ U({\bf{r}})$ the particular solution of (\ref{sch}) can be immediately written down as non-relativistic Volkov states  $\varphi_p({\bf{r}},t)$ which exactly incorporate the interaction with the laser field, 

%%%%%%%%%%%%%%%%%%%%%%%%%%%%%%%%%%%%%%%%
\begin{figure}[!h]
\begin{center}
%* \vspace*{1.0cm} 
%\hspace*{0.5cm}
\scalebox{0.45}{
\rotatebox{0}{\includegraphics{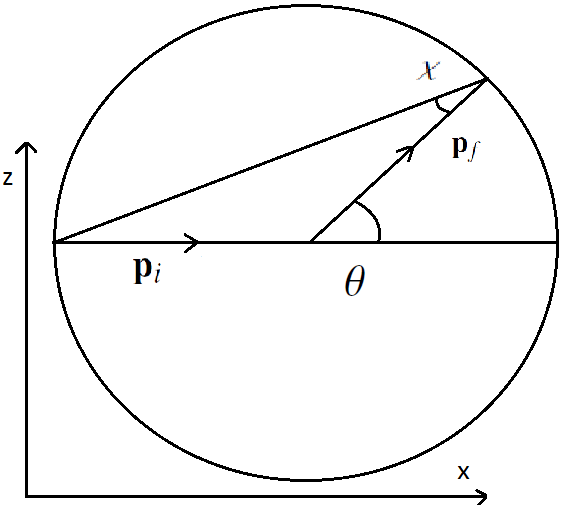}}}
\end{center}
\caption{The geometry of the scattering process. The $^{12}C$  nucleus is in the center of the circle,  ${\bf{p}}_i$ and  ${\bf{p}}_f$ stand for the initial and final scattered proton momenta, $\theta$ is the proton scattering angle, laser pulse propagates parallel to the x axis and linearly polarized in the x-z plain.  The $\chi$ angle is needed for the laser-proton momentum transfer.  }  	
\label{egyes}       % Give a unique label 
\end{figure}
%%%%%%%%%%%%%%%%%%%%%%%%%%%%%%%%%%%%%%%%
%where 
\begin{eqnarray}
%\begin{split}
\varphi_p({\bf{r}},t) = \frac{1}{(2\pi \hbar)^{3/2}} exp\left[    \frac{i}{\hbar}{\bf{p}}\cdot{\bf{r}}  - \right. \nonumber \\
\left.  \int_{t_0}^t dt'  \frac{1}{2m}
\left(  {\bf{p}} - \frac{e}{c} {\bf{A}}(t')  \right)^2 \right]  
%\end{split}
\end{eqnarray}

Volkov states, which are modulated de Broglie waves, parametrized by momenta ${\bf{p}}$ and form an orthonormal and complete set, 
\begin{eqnarray}
\int d^3 r \varphi_p^*({\bf{r}},t) \varphi_{p'}({\bf{r}},t)  = \delta_3 ({\bf{p}}- {\bf{p'}})  \nonumber \\
\int d^3 p \varphi_p({\bf{r}},t) \varphi_{p}^*({\bf{r'}},t)  = \delta_3 ({\bf{r}}- {\bf{r'}}).
\label{norm}
\end{eqnarray}

To solve the original problem of Eq. (\ref{sch}) we write the exact wave function as a superposition of an incoming Volkov state 
and a correction term, which vanishes at the beginning of the interaction (in the remote past $t_0 \rightarrow -\infty$). 
The correction term can also be expressed in terms of the Volkov states, since these form a complete set (see the equation of (\ref{norm})), 

\begin{equation}
\Psi({\bf{r}},t) = \varphi_{p_i}({\bf{r}},t) + \int d^3p a_p(t) \varphi_p ({\bf{r}},t), 
\hspace*{3mm} a_p(t_0) =0. 
\label{nagy}
\end{equation}

It is clear that the unknown expansion coefficients $a_p(t) $ describe the non-trivial transition 
symbolized as ${\bf{p}}_i \rightarrow {\bf{p}}$, from a Volkov state of momentum ${\bf{p}}_i$ 
to another Volkov state with momentum {\bf{p}}. If we take the projection of $\Psi$ into some Volkov state $\varphi_p(t)$ we get 

\begin{equation}
\int d^3 r \varphi_p^*({\bf{r}},t) \Psi({\bf{r}},t)  = \delta_3 ({\bf{p}}- {\bf{p}}_i) + a_p(t). 
\label{norm2}
\end{equation}

Bye inserting $\Psi$ of Eq. (\ref{nagy}) into the complete Schrödinger equation (\ref{sch}), we 
receive the following integro-differential equation for the coefficients $a_p(t)$, 

\begin{eqnarray}
 i\hbar \dot{a}_{p'} (t) = 
\int d^3 r  \varphi_{p'}^*({\bf{r}},t') U({\bf{r}})  \varphi_{p_i}({\bf{r}},t') +  \nonumber \\ 
\int d^3 p  a_p(t) \int d^3 \varphi_{p'}^*({\bf{r}},t') U({\bf{r}})  \varphi_{p}({\bf{r}},t'),  
\end{eqnarray}

where the scalar product was taken with $\varphi_{p'}(t)$ on both sides of the resulting equation 
and the orthogonality property of the Volkov sates was taken after all (see the first Eq. of (\ref{norm})). Owing to the initial condition $a_p(t_0)=0$, displayed already in Eq. (4) the formal 
solution of (6) can be formulated as 

\begin{eqnarray}
 a_{p'} (t) = &
-\frac{i}{\hbar}  \int_{t_0}^t  dt'  \int d^3 r  \varphi_{p'}^*({\bf{r}},t') U({\bf{r}})  \varphi_{p_i}({\bf{r}},t')  \nonumber \\
& -\frac{i}{\hbar}  \int_{t_0}^t  dt'
\int d^3 p  a_p(t') \nonumber \times \\
& \int   d^3 r \varphi_{p'}^* ({\bf{r}},t') U({\bf{r}})  \varphi_{p}({\bf{r}},t').  
\label{inte}
\end{eqnarray}

In the spirit of the iteration procedure used in scattering theory the $(k+1)-$th iterate of $a_p(t)$ is 
expresses by the k-th iterate on the right hand side in  (\ref{inte}) like 

\begin{eqnarray}
 a_{p} ^{(k+1)}(t) =& 
-\frac{i}{\hbar}  \int_{t_0}^t  dt'  \int d^3 r  \varphi_{p'}^*({\bf{r}},t') U({\bf{r}})  \varphi_{p_i}({\bf{r}},t')  \nonumber \\ 
& -\frac{i}{\hbar}  \int_{t_0}^t  dt'
\int d^3 p  a_p^{(k)}(t')  \times \nonumber \\   
& \int d^3 r \varphi_{p'}^*({\bf{r}},t') U({\bf{r}})  \varphi_{p}({\bf{r}},t').  
\label{inte2} 
\end{eqnarray}

%\begin{align}
%\begin{split}
%a_{p}^{(k+1)}(t) =&
%-\frac{i}{\hbar}  \int_{t_0}^t  dt'  \int d^3 r  \varphi_{p'}^*({\bf{r}},t') U({\bf{r}})  %\varphi_{p_i}({\bf{r}},t')  \nonumber \\ 
%& -\frac{i}{\hbar}  \int_{t_0}^t  dt'
%\int d^3 p  a_p^{(k)}(t')  \times \nonumber \\   
%& \qquad \int d^3 r \varphi_{p'}^*({\bf{r}},t') U({\bf{r}})  \varphi_{p}({\bf{r}},t').
%\end{split}
%\end{align}

In the  first Born approximation (where the transition amplitude is linear in the scattering potential $U({\bf{r}})$ ) we receive 
the transition amplitude  in the next form

\begin{eqnarray}
T_{fi} = \lim_{t \rightarrow \infty}\lim_{t_0 \rightarrow -\infty} a^{(1)}_{p_f}(t) =  \nonumber \\
-\frac{i}{\hbar}  \int_{-\infty}^{\infty}  dt'  \int d^3 r  \varphi_{p_f}^*({\bf{r}},t') U({\bf{r}})  \varphi_{p_i}({\bf{r}},t').
\end{eqnarray}
By taking the explicit form of the Volkov states (3) with the bichromatic vector potential  (2)
%\begin{equation}
 %{\bf{A}}(t) =   {\bf{\epsilon_1}} ({\cal{E}}_1/\omega) cos(\omega t) + {\bf{\epsilon}}_m  ({\cal{E}}_m/m\omega)cos(m\omega t + %\tilde{\varphi})
%\end{equation}
note that the $A^2$ term drops out from the transition matrixelement (10), and $T_{fi}$ 
becomes 
\begin{eqnarray}
& T_{fi} = \sum_{n=-\infty}^{\infty} T_{fi}^{(n)} ,  \nonumber \\ 
 &\hspace*{-1cm}T_{fi}^{(n)}  = -2\pi i \delta \left(   \frac{p_f^2-p_i^2}{2m} + n \hbar \omega  \right)
 \frac{ U({\bf{q}})}{(2\pi \hbar)^3}  C_n(a,b;\tilde{\varphi}) ,     
\label{matr}
\end{eqnarray}
%%%
 
where the Dirac delta stand for energy conservation, $U({\bf{r}})$ is the Fourier transformed interaction potential. The main difference to the monochromatic field   
 $C_n(a,b;\varphi)$ which is the generalized Bessel function with the form of 
\begin{equation} 
C_n(a,b; \tilde{\varphi}) = \sum_{\lambda=-\infty}^{\infty} J_{n-m\lambda}(a)J_{\lambda}(b)e^{-i\lambda \tilde{\varphi}}.
\label{gener}
\end{equation}
for the second or third harmonics if $m=2,3$. 
The generalized phase-dependent Bessel functions of the first order can be obtained by expanding its  generating function into a Fourier series viz. 
\begin{equation}
e^{ \left\{ i [ a sin(\omega t) + b sin(n\omega + \tilde{{\varphi}}) ] \right\} } =  
 \sum_{n=-\infty}^{\infty}    e^{i n \omega t }  C_n(a,b;\tilde{\varphi}). 
\end{equation}
Here the various different Fourier components of the expansions of the two exponential 
on the left-hand side into ordinary Bessel functions $J_{\lambda}$ yielding the same final 
harmonic frequency  $n\omega$ may be considered in this classical problem as the infinite 
number of different phase-dependent "{\it{reaction channels}}" contributing to the matrix element 
 of (\ref{matr}).  
Note that this is a generalization of the Jacobi-Anger formula which was used in the 
monochromatic case  
\begin{equation}
e^{i a sin(\omega t)} = \sum_{\lambda=-\infty}^{\infty} J_{\lambda}(a)e^{i \lambda \omega t}.  
\end{equation}
With the well-known symmetry property of  the Bessel functions of the first kind $J_{-\lambda}(a) =  (-1)^{\lambda} J_{\lambda}(a)  $  we can easily verify the following symmetry relations for the second and third harmonics (\ref{gener})  
\begin{eqnarray}
C_{-n}(a,b_2;\tilde{\varphi})&=& (-1)^n C_n^*(a,b_2;{\tilde\varphi}-\pi), \nonumber \\ 
C_{-n}(a,b_3;\tilde{\varphi})&=& (-1)^n C_n^*(a,b_3;\tilde{\varphi}).   
\end{eqnarray}

%%%%%%%%%%%%%%%%%%%%%%%%%%%%%%%%%%%%%%%%%%%%%%%%%%%%%%%%%

The  $U({\bf{q}})$ is the Fourier transformed of the scattering potential with the momentum transfer of 
 ${\bf{q}}  \equiv {\bf{p}}_i  - {\bf{p}}_f  $ where  ${\bf{p}}_i$ is the initial and 
${\bf{p}}_f$ is the final proton momenta.    The absolute value is $    q = \sqrt{p_i^2 + p_f^2 - 2p_ip_f cos(\theta_{p_i,p_f})}$. 
In our case, for 49 MeV energy protons absorbing optical photons the following approximation is valid 
$q \approx 2 p_i  |sin(\theta/2)| $.
As we mentioned earlier the polarization vector of the laser pulses are parallel with the inital proton momenta 
$\epsilon \parallel {\bf{p}}_i$, then ${\bf{\cal{\epsilon}}} ({\bf{p}}_i -{\bf{p}}_f )  = p_i (1-cos(\theta)) $. 

 The Dirac delta describes photon absorptions $(n<0)$ and emissions $(n>0)$ with energy conservation.  
The arguments of the two Bessel functions are the following
\begin{equation}
a = \frac{m_e}{m_p}a_0  {\bf{\cal{\epsilon}}} ({\bf{p}}_i -{\bf{p}}_f )c/{\hbar \omega}
\end{equation}
\begin{equation}
b_m = \frac{m_e}{m_p}a_m  {\bf{\cal{\epsilon}}}_m ({\bf{p}}_i -{\bf{p}}_f )c/{\hbar m \omega_0}
\end{equation}
 where $a_m = \frac{e E_m}{m_e c m\omega} = 10^{-9} \sqrt{I}_m/E_{ph}$. 
where the laser energy $\hbar \omega$ is measured in eV, the proton energy $E_p$  in MeV and the laser intensity I in W/cm$^2$. 

Collecting the constants together
the very final formulas for the arguments are 
\begin{equation}
a = 10^{-4} \sqrt{I} [1-cos(\theta)],  \hspace*{5mm}
b_m = \frac{10^{-4} \sqrt{I}_m [1-cos(\theta)]}{m}. 
\end{equation} 
(Note, that this formula is valid for any kind of external laser field. For a 49 MeV proton projectile even the 10 keV X-ray laser has a negligible energy.) 
%%%%%%%%%%%%%%%%%%%%%%%%%%%%%%%%%%%%%%%%%
 %%%%%%%%%%%%%%%%%%%%%%%%%%%%%%%%
The final differential cross section formula for the laser associated collision with simultaneous n$^{th}$-order photon absorption and emission processes is a bichromatic field of a frequency plus 
it's  m$^{th}$-order high-harmonic  is 
\begin{equation}
\frac{d \sigma^{(n,m)}}{d \Omega} = \frac{p_f}{p_i} \frac{d \sigma_B}{d \Omega} 
\left|  \sum_{\lambda = -\infty}^{\infty} J_{n-m\lambda}(a)J_{\lambda}(b_m)e^{-i\lambda \tilde{\varphi}}  \right|^2
\label{cross}
\end{equation} 
where  $ \frac{d \sigma_B}{d \Omega} $
is the usual Born cross section for the scattering on the potential as was mentioned above. 
At the low energy limit, where the photon energy is much below the kinetic energy of the scattered particle 
the Born cross section can be extracted from the sum. 
The expression Eq. (\ref{cross}) was first calculated by \cite{varro1,varro2,varro3}. 

If we investigate $d\sigma^{n,m}/d\sigma_B = | C_n(a,b_m,\tilde{\varphi})  |^2$, we can study the modification of the cross sections 
due to the interaction with the laser fields. This is the main goal of the recent study.

%\begin{equation}
%\frac{d \sigma^{(n)}}{d \Omega} = \frac{p_f}{p_i}J_n^2(z) \frac{d \sigma_B}{d \Omega}. 
%\label{cross}
%\end{equation} 
%The $ \frac{d \sigma_B}{d \Omega} = \left(\frac{m}{2\pi \hbar^2} \right)^2 |U({\bf{q}})|^2$ 
%is the usual Born cross section for the scattering on the potential $ U({\bf{r}})$ alone (without the laser field). 
%The expression Eq. (\ref{cross}) was calculated with different authors using different methods \cite{bunk, %bunk1,faisal,gont,bergou,bergou2,kroll,faisal2}. \\ 

We must say some words about the central scattering potential $U(r)$ which is 
 the sum of the Coulomb potential of a uniform charged sphere \cite{rudchik} and a short range optical  \cite{woods} potential 
\begin{equation}
U(r) = V_c(r) + V_{ws}(r) + i[W(r) + W_s(r)] + V_{ls}(r) {\bf{l}} \cdot {\bf{\sigma}}
\label{teljes} 
\end{equation} 
where the Coulomb  term is 
\begin{eqnarray}
V_c &=& \frac{Z_p Z_t e^2}{2R_0} \left(3 - \frac{r^2}{R_c^2} \right) \hspace*{0.8cm} r < R_c  \nonumber  \\
V_c &=&\frac{Z_pZ_t e^2}{r}  \hspace*{2.2cm} r \ge R_c  
\end{eqnarray}
where $R_c = r_0 A_t^{1/3}$ is the target radius calculated from the mass number of the  target 
with $r_0 = 1.25 fm$.   $Z_p, Z_t$ are  the charge of the projectile and the target and $e$ is the elementary charge. 
This kind of regularized Coulomb potential helps us to avoid singular cross sections and routinely used in nuclear physics. 

The short range nuclear part  is given via 
\begin{eqnarray}
V_{ws}(r) &=& - V_r f_{ws}(r,R_0,a_0) \nonumber \\ 
W(r) &=& - V_v f_{ws}(r,R_s,a_s) \nonumber \\ 
W_s(r) &=& - W_s(-4a_s) f_{ws}'(r,R_s,a_s) \nonumber \\
 V_{ls}(r) &=& -(V_{so} + iW_{so})(-2) g_{ws}(r,R_{so},a_{so}) \nonumber \\ 
f_{ws}(r,R,a) &=& \frac{1}{1+ exp \left( \frac{r-R}{a} \right)} \nonumber \\ 
	f'_{ws} (r,R,a)&=& \frac{d}{dr} f_{ws}(r,R,a) \nonumber \\   
g_{ws}(r,R,a) &=& f'_{ws}(r,R,a)/r.
\end{eqnarray}
The constants $V_r, W_v,V_{so} $ and $ W_{so}$ are the strength parameters, 
and $a_{0,s,so}, R_{0,s,so}$ are the diffuseness and the radius parameters given for large number of nuclei. The $f$ function is called the shape function of the interaction. 
As we will see at moderate collisions energies the complex terms become zero. 
According to the work of  \cite{hlope} the complete analytic form of the Fourier transform of the WS potential can be calculated 
on the complex plain with contour integration using the residuum theorem. 
%via the  following kind of complex integrals 
 %$V(q) =  \int_0^{\infty} dz \frac{z \exp(i\rho_k z)}{1 + \exp(z-\alpha_k)}, $ where 
%$\rho_k =qa_k$, $\alpha_k =R_k/a_k$ and $z = r/a_k$ are dimensionless variables. 
% The integrals can be evaluated by contour integration using the residuum theorem. 
 For exhaustive
details see  \cite{hlope}.  
The Fourier transformed second term of Eq. (\ref{teljes}) reads 
\begin{eqnarray} 
\label{ws}
& \hspace*{-1cm} V_{ws}(q) =  \frac{V_r}{\pi^2} \left\{ \frac{\pi a_0 e^{-\pi a_0 q}}{q(1-e^{-2\pi a_0 q})^2} \left[ 
R_0(1-e^{-2\pi a_0 q})cos(qR_0) -  \right. \right.    \nonumber \\ & \hspace*{-2cm}  \left.\pi a_0 (1+ e^{-2\pi a_0 q}) sin(qR_0) \right]
- \nonumber \\ 
 & \hspace*{-2cm}\left. a_0^3 e^{-\frac{R_0}{a_0}} \left[ \frac{1}{(1+a_0^2q^2)^2} - 
\frac{2e^{-\frac{R_0}{a_0}}}{(4+
a_0^2 q^2)^2}  \right]    \right\} .  
\end{eqnarray}
%For the $W(q)$ imaginary term, the same expression was derived with $W_v,a_s,R_s$ instead of $V_r,a_0$ and $R_0$.
The surface term $W_s(r)$  (fourth term in Eq. (\ref{teljes})) gives the following formula in the momentum space: 
\begin{eqnarray}
\label{s}
&\hspace*{-1.5cm} W_s(q) = - 4 a_s \frac{W_s}{\pi^2} \left\{ \frac{\pi a_s e^{-\pi a_s q}}{(1-e^{-2\pi a_s q})^2} \left[ 
(\pi a_s(1+e^{-2\pi a_s q}) \right.   \right.  - \nonumber \\  
&\hspace*{-1cm}  \left.  \frac{1}{q} (1- e^{-2\pi a_s q}) )cos(qR_s)   + R_s(1-e^{-2\pi a_s q})sin(qR_s)   \right] +  \nonumber \\ 
&\hspace*{-3.5cm}\left.   a_s^2 e^{-\frac{R_s}{a_s}} \left[ \frac{1}{(1+a_s^2q^2)^2} - \frac{4e^{-\frac{R_s}{a_s}}}{(4+
a_s^2 q^2)^2}  \right]    \right\}.    
\end{eqnarray}
 The last term in Eq. (\ref{teljes}), the transformed spin-orbit coupling term leads to  
\begin{eqnarray}
\label{ls}
&\hspace*{-1cm}V_{ls}(q) =    -\frac{a_{so}}{\pi^2}(V_{so}+ iW_{so})
 \left\{   \frac{2 \pi e^{-\pi a_{so}q}}{1-e^{-2\pi a_{so}q}} sin(q R_{so})  + 
\right. \\ \nonumber
%\left.
&\hspace*{-1cm} \left.
e^{\frac{-R_{so}}{a_{so}}} \left(  \frac{1}{1+a_{so}^2 q^2} -   \frac{2e^{-R_{so}/a_{so}} {}}{4+a_{so}^2 q^2}\right)
  \right\}.
\end{eqnarray}
where ${\bf{q}} $ is the momentum transfer as defined above.
%is defined  as above  $ {\bf{q}} \equiv  {\bf{p_i}} - {\bf{p_f}}. $ 
The  low energy transfer approximation formula $q \approx 2 p_i  |sin(\theta/2)| $ is valid. 
%%%%
The Fourier transform of the charged sphere Coulomb field is also far from being trivial
 \begin{eqnarray}
&\hspace*{-2.7cm} V_c(q) =    \frac{Z_pZ_t e^2}{2^{\frac{5}{6}} \sqrt{\pi} q^3} \left( -2\cdot 3^{\frac{1}{3}} q  \cos[2^{\frac{2}{3}} 3^{\frac{1}{3}} q]  +  \right.   \nonumber\\ 
&\hspace*{-1cm}  \left.2^{\frac{1}{3}} (1 + 2\cdot 2^{\frac{1}{3}} 3^{\frac{2}{3}} q^2) \sin[
   2^{\frac{2}{3}} 3^{\frac{1}{3}} q] \right)  +     \nonumber  \\
 &\hspace{-1cm}3Z_pZ_t e^2 \sqrt{\frac{2}{\pi}} \left(   \frac{i \pi|q|  }{2q}  - \mathrm{Ci} [2^{\frac{2}{3}}
  3^{\frac{1}{3}} q] + \log(q) - \log|q| - \right. \nonumber \\
&\hspace*{-3cm} \left.  i\mathrm{Si} [2^{\frac{2}{3}}  3^{\frac{1}{3}} q] \right)   
\end{eqnarray}
\label{vc}
 where Ci and Si  are the cosine integral and the sine integral functions for details see \cite{abr}, respectively. 

%%%%%%%%%%%%%%%%%%%%%%%%%%%%%%%%%%%%%%%%%%%%%%%%%%%%%%%%%%%%
\section{Results} 

We applied the outlined method to 49 MeV proton - $^{12}$C scattering.
The parameters of the Woods-Saxon potential are the following the three potential strength $V_R, W_s,V_{so}$ are $ 31.31, 5.98, 2.79$ MeV, the three diffuseness $a_0,a_s, a_{so}$ are $0.68,0.586,0.22$ fm and the three radius parameters $R_0,R_s,R_{so}$
are $1.276,0.89,0.716 $ fm. 
 
%The shape of the 
%Table I contains the parameters of the used Woods-Saxon potential according to the work of \cite{abd}.  

The shape of the differential cross section, which is proportional to the Fourier transformed of the various Woods-Saxon potential  terms for 49 MeV elastic proton -  $^{12}$C scattering Eq.(\ref{ws},\ref{s},\ref{ls}) are presented and analyzed in our former study \cite{barna2} in details. Our calculated total cross section of the elastic scattering is  201 mbarn which is
consistent with the data of \cite{abd}.    

\begin{figure}[!h]
%* \vspace*{1.0cm} 
\scalebox{0.38}{
\rotatebox{0}{\includegraphics{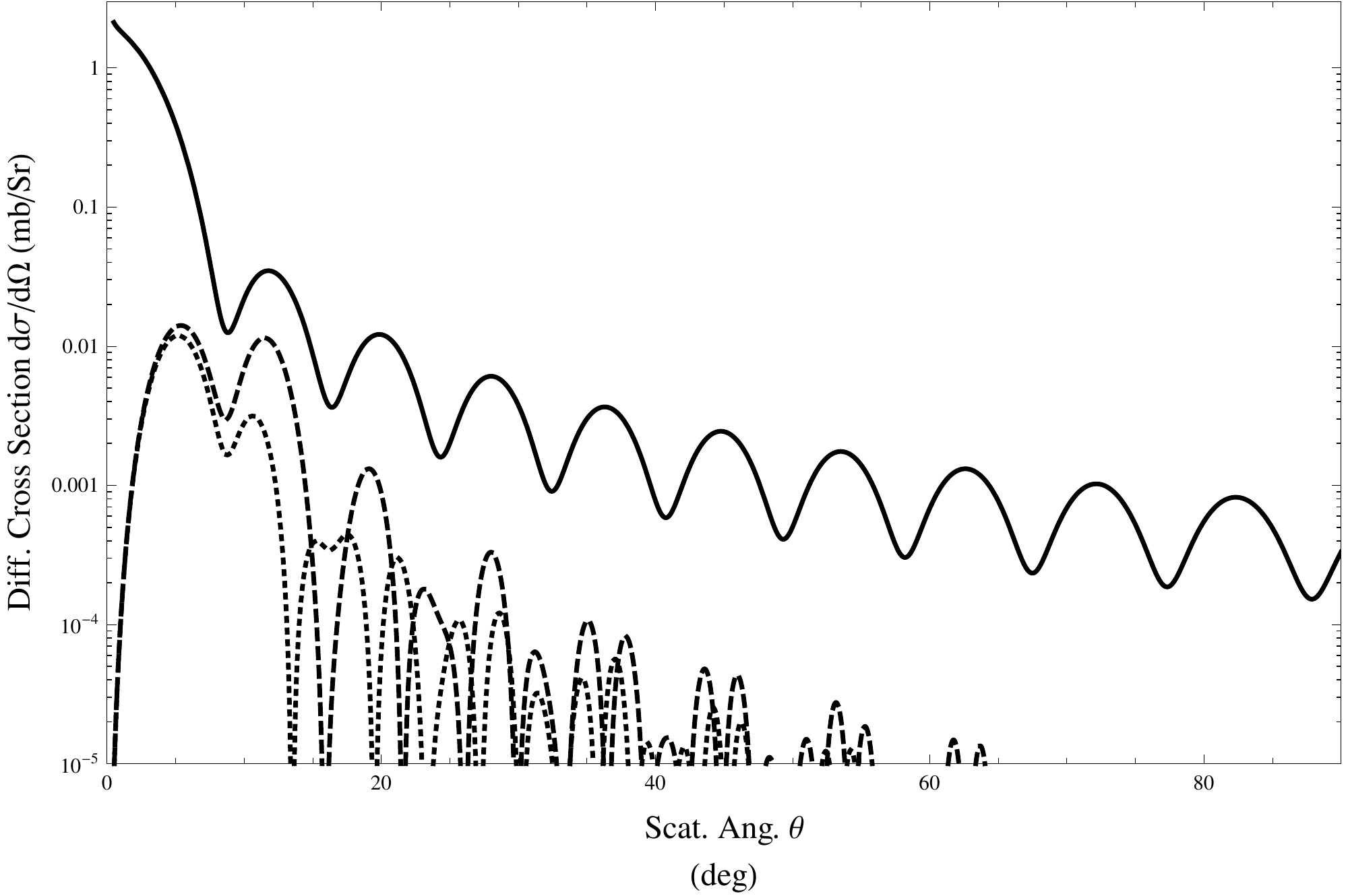}}}  
\caption{  The calculated angular differential cross sections  for $I =  10^{12}$  W/cm$^2 $ laser field intensity. 
The thick solid line is the $n=0$ elastic Born cross section, the dashed line is for the monochromatic case for one photon absorption, and the doted line is for bichromatic fields at single photon absorption as well 
 Eq. (19).   The intensity ratio is $I/I_2 = 2$  with $\tilde{\varphi} = 0$  relative phase difference.}    	
\label{kettes}       % Give a unique label 
\end{figure}

Figure \ref{kettes} shows  the angular differential cross section for the elastic, monochromatic and bichromatic fields at 
$10^{12}$ $W/{cm^2}$ laser intensities. In our former study we found that at higher laser intensities the laser assisted cross sections 
are many magnitudes below the elastic Born cross sections. 
This can be understood from the mathematical properties of the Bessel functions. The maximum (with respect to the index $n$) 
of the Bessel function is around $n_{max} \approx z $  where $z$ is the argument.
If the intensity is $I = 10^{16}$ $W/cm^2$ the argument of (18) is about $10^4$. 
This also means that for a satisfactory convergence to calculate the generalized Bessel functions the value of the sum $\lambda = 1000 $ should be.  For such large values of the argument 
$( |n| << |z| )$ the following asymptotic expansion can approximate \cite{rizs} (formula 8.451)
\begin{equation}
J_{\pm n}(z) \approx \sqrt{\frac{2}{\pi z}}cos(z \mp n\pi/2 -\pi/4).
\end{equation}
For small arguments however, $(|z| << |n| )$ the power expansion is valid $J_n(z) \approx (z/2)^n$  \cite{rizs} (formula 8.440). 
As we mentioned earlier instead of the complete angular differential cross section it is enough to investigate the properties of 
$d \sigma_{Total} / d\sigma_{Born} = J^2_n(z) $  for monochromatic or  $d \sigma_{Total} / d\sigma_{Born} = |C_n(a,b_m,\tilde{\varphi})|  ^2  $ for bichromatic laser assisted scattering to know how the  differential cross sections behave at parameter changes. 
Is is known that $ \sum_{\lambda=-\infty}^{+\infty} J_{\lambda}^2(z) = 1$ or similarly $  \sum_{\lambda=-\infty}^{+\infty} 
 |C_{\lambda}(a,b_m,\tilde{\varphi})|  ^2 = 1. $ Therefore the total inelastic contributions (where any photons are absorbed or emitted) are the following $1 - J_0^2(z)$ for the monochromatic case and $1 -|C_0(a,b_m,\tilde{\varphi})| ^2 $ for bichromatic 
case. Figure 3 presents such results for $I = 10^{12}$ $W/cm^2$ intensities. Note, the remarkable contributions at small scattering angles. 

The role of the relative phase between the two corresponding frequencies $\tilde{\varphi}$, as a coherent control parameter is also 
instructive to examine. First we may fix the scattering angle to a fix degree say $\theta = 7^{\circ}$  and investigate how the various 
phases change the absolute value of the  $C_n$. 
Figure 4 presents the relative phase dependence of $|C_n(a,b_2,\tilde{\varphi})| $  for one,two and three photon absorptions. 
The major contribution is at the single photon absorption which meets our physical intuition. 

Figure 5 presents the  role of the intensity rations $I/I_2$ at the above fixed ange for various relative phase $ \tilde{\varphi}$. 
The larger the intensity of the second harmonic the  larger the role of the relative phase.

%%%%%%%%%%%%%%%%%%%%%%%%%%%%%%%%%%%%%%%%% 
%%%%%%%%%%%%%%%%%%%%%%%%%%%%%%%%%%%%%%%%%%%%%%%%%%%%
%%%%%%%%%%%%%%%%%%%%%%%%%%%%%%%%%%%%%%%%%%%%%%%%%%%
\begin{figure}[!h]
%* \vspace*{1.0cm} 
%\hspace*{0.5cm}
\scalebox{0.38}{
\rotatebox{90}{\includegraphics{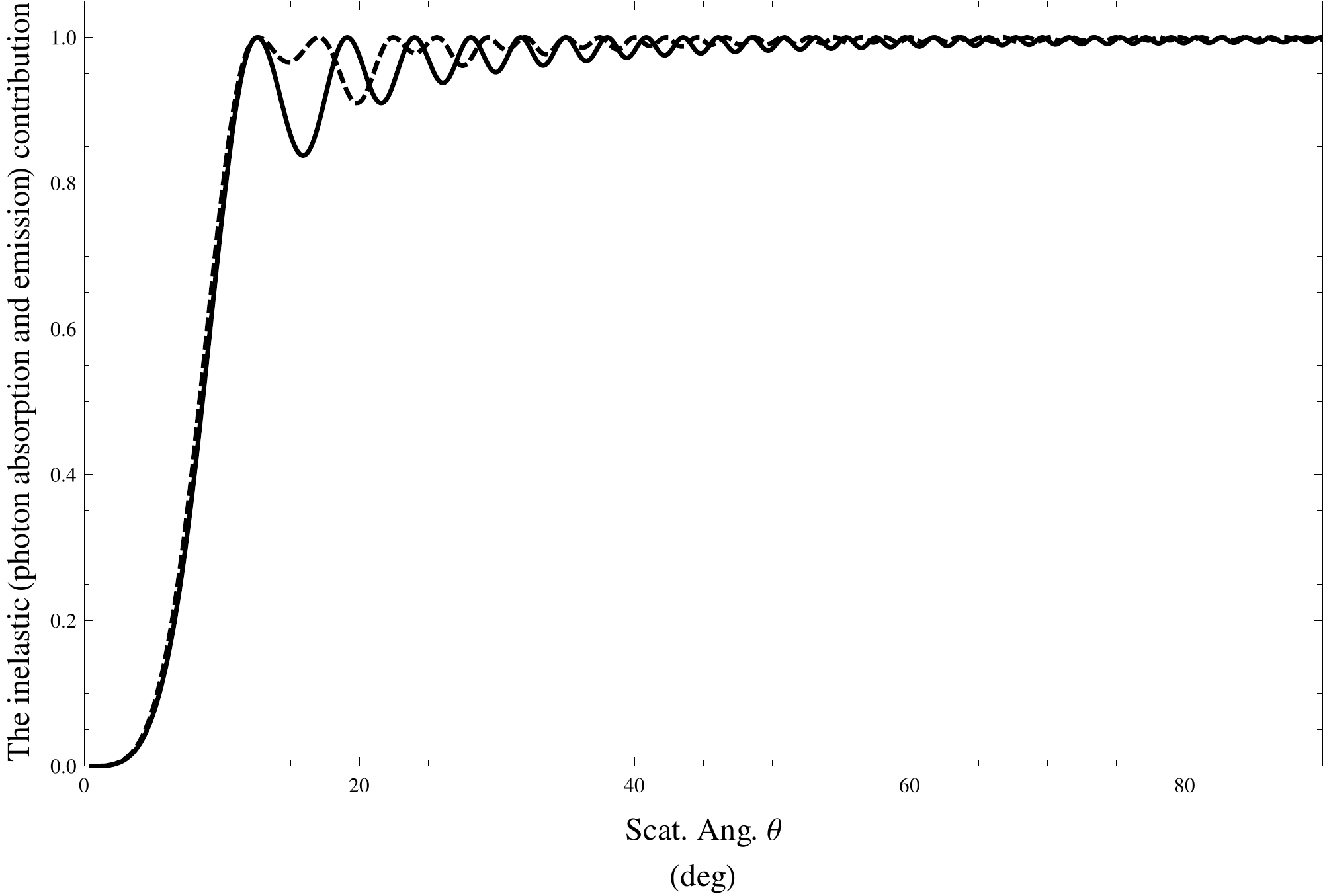}}}  
\caption{The total inelastic contributions of the laser assisted scattering. Solid line is for the monocromatic and the dashed line is for the bichromatic field.  Parameters are as the same as in Fig. 2. }    	
\label{harmas}       % Give a unique label 
\end{figure} 
%%%%%%%%%%%%%%%%%%%%%%%%%%%%%%%%%%%%%%%%%%%%%% 
 \begin{figure}[!h]
%* \vspace*{1.0cm} 
%\hspace*{0.5cm}
\scalebox{0.38}{
\rotatebox{0}{\includegraphics{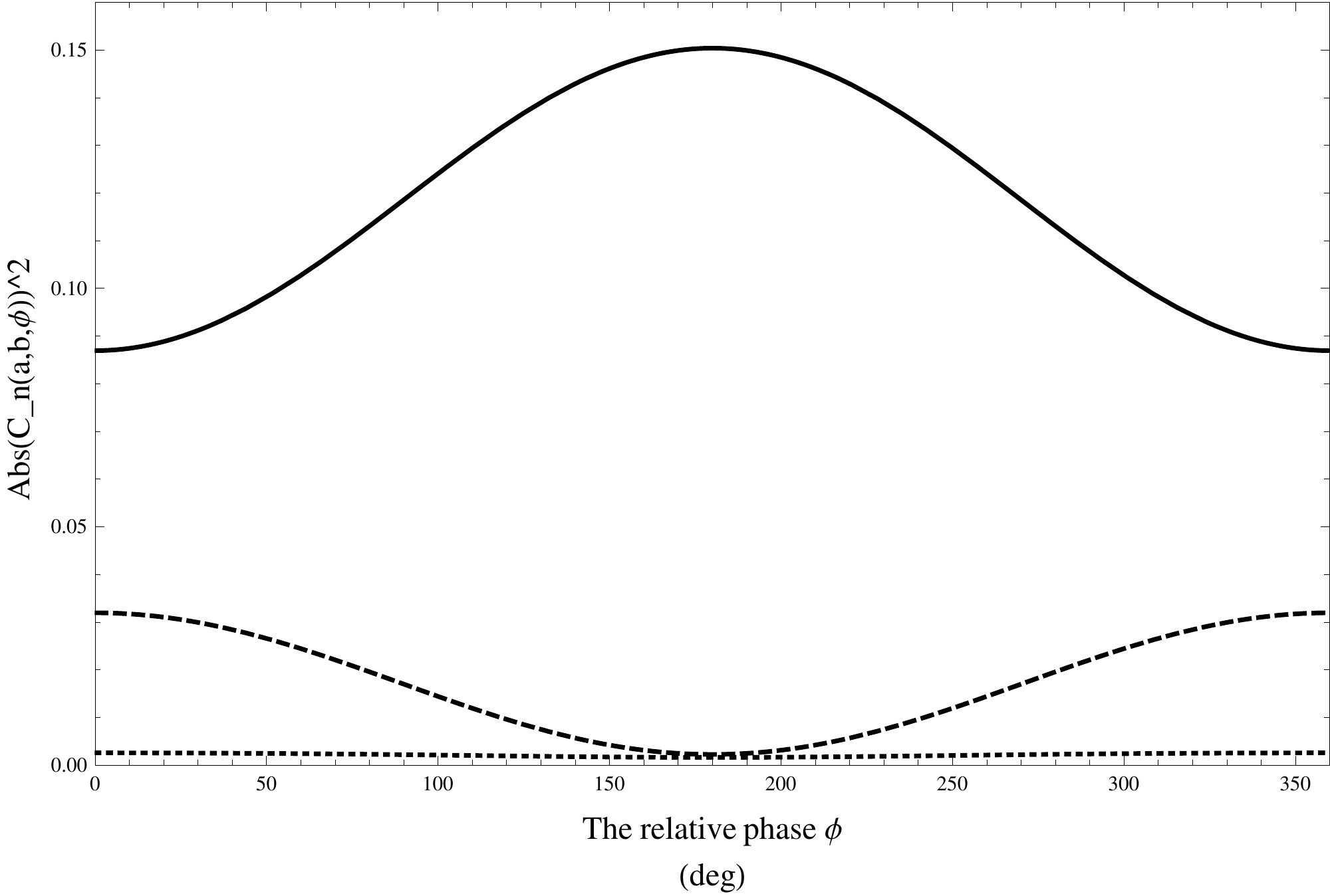}}}  
\caption{Shows the dependence on the relative phase between the fundamental and the second harmonics of the same intensity $I=10^{12}$ $W/cm^2.$ 
The solid line is for one photon the dashed line is for two photon and the dotted line is for three photon absorptions.}    	
\label{negyes}       % Give a unique label 
\end{figure} 
%%%%%%%%%%%%%%%%%%%%%%%%%%%%%%%%%%%%%%%%%%%%
 \begin{figure}[!h]
%* \vspace*{1.0cm} 
%\hspace*{0.5cm}
\scalebox{0.38}{
\rotatebox{0}{\includegraphics{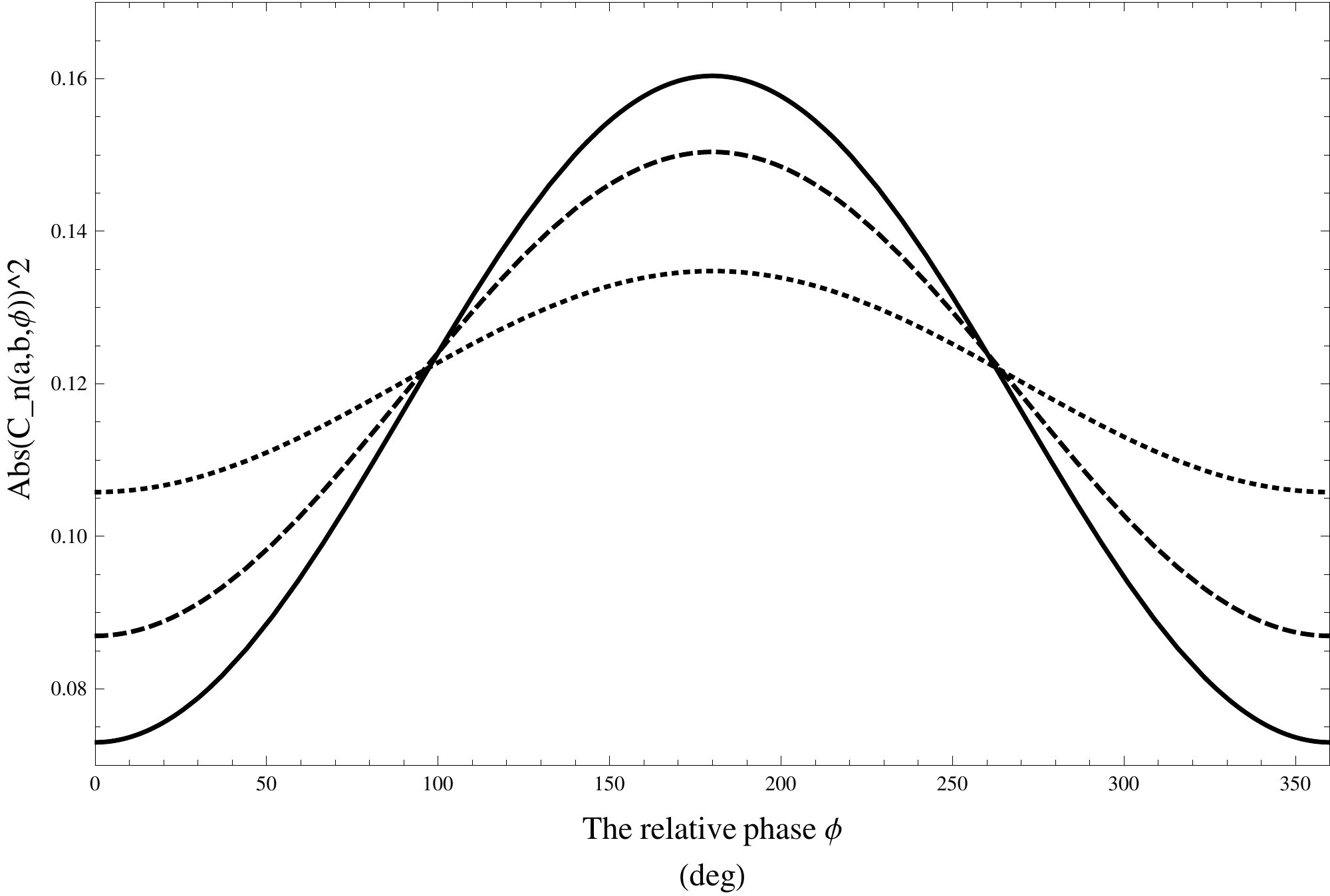}}}  
\caption{The role of relative intensity of the  two frequencies $I/I_2$ for single photon absorption as the function of the relative phase $\tilde{\varphi}$. 
The solid line refers to the numerical value  $I/I_2=1$, the dashed line to 2 and the dotted line to 10 intensity ratios, respectively.}    	
\label{otos}       % Give a unique label 
\end{figure} 

\section{Summary}

We presented a formalism which gives an analytic angular differential cross section formula for proton nucleon scattering on a Woods-Saxon optical potential in a bichromatic laser field where the n$^{th}$-order photon absorption end emission is taken into account simultaneously in a non-perturbative way. 
As a physically relevant example we investigated the proton - $^{12}$C collision system at moderate 49 MeV proton energies in the field of an optical Ti:sapphire system with wavelength of 800 nm and its second harmonic. 
The role of the relative phase between the fields are investigated. 
%in the  range of   $10^{11}$ W/cm$^2$ to $  10^{21}$ W/cm$^2.$  
%As a second system we took a 10 keV X-ray laser field with $  10^{16}$ W/cm$^2$ intensity.   
 %The calculated cross sections are much lower than the elastic cross sections in all cases. 
We hope that our study will give a strong impetus and bring the nuclear and laser physics community together to perform such experiments in the ELI or X-FEL facilities which will be available in a couple of years. 

\section{Acknowledgment}     
S. V. has been supported by the National Scientific Research Foundation OTKA, Grant No. K 104260.
 Partial support by the ELI-ALPS project is also acknowledged. The ELI-ALPS project (GOP-1.1.1-12/B-2012-0001) is supported by the European Union and co-financed by the European Regional Development Fund. 

%%%%%%%%%%%%%%%%%%%%%%%%%%%%%%%%%%%%%%%%%%%%%%%%%%%%%%%%%%%%

%% main text
\section{References}
\label{}

%% The Appendices part is started with the command \appendix;
%% appendix sections are then done as normal sections
%% \appendix

%% \section{}
%% \label{}

%% If you have bibdatabase file and want bibtex to generate the
%% bibitems, please use
%%
%%  \bibliographystyle{elsarticle-num} 
%%  \bibliography{<your bibdatabase>}

%% else use the following coding to input the bibitems directly in the
%% TeX file.

\bibliographystyle{elsarticle-num}
\bibliography{bv_refs}
\end{document}